\documentclass[10pt,twosides,a4paper]{book}
\usepackage{physics} 
\usepackage{amssymb}
\usepackage[pdftex]{graphicx}
\usepackage{latexsym}
\usepackage[small]{caption2}
\usepackage{amsmath}
\usepackage[english]{babel}
\usepackage{babel}
\usepackage{amsmath,amsthm}
\usepackage{rotating}
\usepackage{multirow}
\usepackage{graphicx}
\usepackage{tocloft}
\usepackage{bbm}
\usepackage{dsfont}
\usepackage{amsfonts}
\usepackage{accents}
\usepackage{pdfpages}
\usepackage[thinc]{esdiff}
\usepackage{hyperref}
\usepackage[numbers,sort&compress,square]{natbib}
\usepackage{natbib}
\usepackage{pdfsync}
\usepackage{fancyhdr}
\usepackage{scalerel,stackengine}
\usepackage{fancyhdr}
\renewcommand{\cftchappresnum}{CHAPTER }
\newlength{\mylen}
\fancypagestyle{plain}{}

\begin{document}

\chapter*{Relativistic approach to thermodynamic irreversibility}

\markboth{Relativistic approach to thermodynamic irreversibility}{M. Basso, J. Maziero and L. C. Céleri}

\noindent{\large \textbf{M. Basso$^{1}$, J. Maziero$^{2}$ and L. C. Céleri$^{3,a}$}}

\vspace{3mm}

\noindent{$^1$ Center for Natural and Human Sciences, Federal University of ABC, Avenue of the States, Santo Andr\'e, S\~ao Paulo, 09210-580, Brazil
\\ $^2$ Department of Physics, Center for Natural and Exact Sciences, Federal University of Santa Maria, Roraima Avenue 1000, Santa Maria, Rio Grande do Sul, 97105-900, Brazil
\\ $^3$ QPequi Group, Institute of Physics, Federal University of Goi\'as, Goi\^ania, Goi\'as, Brazil}


\noindent{\texttt{$^{a}$lucas@qpequi.com}}

\section{Introduction}

Irreversible processes are ubiquitous in Nature despite the fact that the fundamental laws of Nature respect the time-reverse symmetry~\cite{zeh}. Irreversibility is signaled by thermodynamic entropy production~\cite{landi} and says that the thermodynamic arrow of time points from low to high entropy~\cite{parrondo}. One of the fundamental results in this context is the set of relations known as fluctuation theorems, which state that in every physical process, a positive entropy production is highly likely to be observed, while processes in which entropy decreases are observed with vanishingly small probability~\cite{Crooks1998,Jarzynski2011,Seifert2012,Esposito2009,Campisi}. Therefore, an average positive entropy production is typically manifested in Nature. 

The first ideas on relativistic thermodynamics were put forward by Einstein and Planck while considering the behavior of thermodynamic properties under changes in reference frames~\cite{einstein1907,einstein1989,planck1908}. An important step in the direction to understand the connections between these theories was the development of black hole thermodynamics~\cite{Wald1994}. Such laws were then employed to show that Einstein's field equations could be interpreted as a thermodynamic equation of state~\cite{Jacobson1995}. Many advances in this direction were witnessed, among which we mention investigations about the non-equilibrium properties of spacetime~\cite{Eling2006}, attempts to construct a statistical mechanical theory of the gravitational field, and the conjecture that time can have a thermodynamic origin~\cite{Rovelli1993,Connes1994, Rovelli2011,Rovelli2013}.

In this chapter, we describe a major step in the direction to properly understand the thermodynamics of localized quantum systems living in a general curved spacetime and, thus, under the action of the gravitational field~\cite{Basso2023,Basso2024}. Considering linear response, a fluctuation theorem in curved spacetimes was presented in Ref.~\cite{Mottola1986}, while the non-equilibrium fluctuations of a black hole horizon were studied in Ref.~\cite{Iso2011} in the context of the Jarzynski equality~\cite{Jarzynski1997} along with the generalized second law of thermodynamics~\cite{Bekenstein1974}. A fluctuation theorem for a quantum field in a specific model of an expanding Universe was described in Ref.~\cite{Liu}.

This chapter describes a fully general relativistic detailed quantum fluctuation relation for a localized quantum system~\cite{Basso2023,Basso2024}. The result is based on the so-called two-point measurement (TPM) scheme~\cite{Esposito2009}. Therefore, we unveil the complete impact of the gravitational field on irreversible processes by explicitly considering the effect of spacetime curvature. One of the most impressive consequences of this result is that thermodynamic entropy is fundamentally observer-dependent, being deeply linked to the time-orientability of the spacetime. We present two paradigmatic examples in which we discuss the role of the equivalence principle and the expansion of the universe on entropy generation.

We use natural units throughout the paper. The signature of the metric is $(-,+,+,+)$ and $\eta_{a b} = \text{diag}(-1,1,1,1)$ is the Minkowski metric.

\section{Localized quantum systems in a curved spacetime}
\label{sec:ii}

Our initial goal is to formulate a description of a localized non-relativistic quantum system in a curved spacetime. This system will be characterized by a set of external degrees of freedom along with one or more internal degrees of freedom. To achieve this, we construct Fermi normal coordinates around a timelike trajectory representing the worldline of our laboratory frame. At last, we analyze the Hamiltonian dynamics of a localized quantum particle in the vicinity of this trajectory~\cite{Poisson2011,Perche2022}.

\subsection{Fermi normal coordinates}
\label{sec:iia}
To begin with, let $(\mathcal{M}, \mathbf{g})$ be a four-dimensional spacetime, where $\mathcal{M}$ is a differentiable manifold and $\mathbf{g}$ is a Lorentzian metric. In this spacetime, the worldline of a laboratory frame is represented by a timelike curve $\gamma: I \subset \mathbb{R} \to \mathcal{M}$, parametrized by its proper time $\tau \in I$, with 4-velocity $u^{\mu}$ which satisfies $u_{\mu} u^{\mu} = -1$.

\begin{figure}
    \centering
    \includegraphics[scale=0.5]{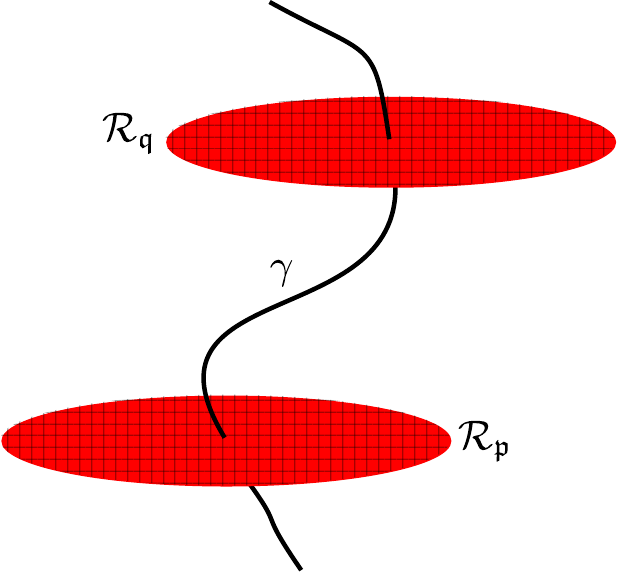}
    \caption{The laboratory world-line $\gamma$ with $4$-velocity $u^{\mu}$ and the rest spaces $\mathcal{R}_{\mathfrak{p}}$ and $\mathcal{R}_{\mathfrak{q}}$ where the Fermi normal coordinates are valid.}
    \label{fig:fermi}
\end{figure}

In order to define the Fermi normal coordinates, we begin by setting the proper time $\tau$ along the curve $\gamma$ as the time component of the coordinate system, as illustrated in Fig.~\ref{fig:fermi}. Next, we define an orthonormal basis $e_{a}^{\ \mu}$ at a point $\mathfrak{p} \equiv \gamma(\tau = 0) \in \mathcal{M}$, where $a = 0, 1, 2, 3$ labels the four basis vectors, with $e_0^{\ \mu}$ identified as the tangent vector $u^{\mu}$. At point $\mathfrak{p}$, we have the relation
\begin{align}
    g_{\mu \nu} e_{a}^{\ \mu} e_{b}^{\ \nu} =  \eta_{a b},
\end{align}
where $\eta_{a b} = \text{diag}(-1,1,1,1)$ is the Minkowski metric. The next step is to extend this orthonormal frame along the curve $\gamma$ while preserving the orthogonality of the basis vectors. Parallel transport along $\gamma$ alone does not ensure the vectors in the set $\{e_{a}^{\ \mu}\}_{a = 0}^3$ will remain mutually orthogonal, and tangent to $\gamma$, unless $\gamma$ is a geodesic. To maintain the orthonormality of the frame along $\gamma$, we employ Fermi-Walker transport, governed by the differential equations~\cite{hawkingellis}
\begin{align}
    \frac{\textrm{D}_F}{\dd \tau} (e_{a})^{\mu} \equiv \frac{\textrm{D}}{\dd \tau} (e_{a})^{\mu} + 2 a^{[\mu}u^{\nu]}(e_a)_{\nu} = 0, \label{eq:fermit}
\end{align}
where $\textrm{D}/\dd \tau \equiv u^{\mu} \nabla_{\mu}$ is the covariant derivative along the curve $\gamma$, $a^{\mu} = (\textrm{D}/\dd \tau)\, u^{\mu} = u^{\nu} \nabla_{\nu} u^{\mu}$ is the $4$-acceleration of the curve $\gamma$ while $2 a^{[\mu}u^{\nu]} \equiv a^{\mu} u^{\nu} - a^{\nu} u^{\mu}$. When the above condition is met, the vectors $e_{a}^{\ \mu}$ are said to be Fermi transported. We refer to this frame as the laboratory frame.

To define the spacelike Fermi normal coordinates $(x^1, x^2, x^3)$, we first consider the normal neighborhood of the point $\mathfrak{p} = \gamma(\tau = 0)$, which consists of all points that can be reached from $\mathfrak{p}$ by a single geodesic. This neighborhood is denoted by $\mathcal{U}_{\mathfrak{p}}$. Within this region, we define the local rest space $\mathcal{R}_{\mathfrak{p}} \subset \mathcal{U}_{\mathfrak{p}}$ as the set of points connected to $\mathfrak{p}$ by geodesics whose tangent vectors are orthogonal to $u^{\mu}$, the 4-velocity at $\mathfrak{p}$. Refer to Fig.~\ref{fig:fermi} for an illustration. 

With this setup, we can assign the coordinates $(\tau = 0, x^1, x^2, x^3)$ to any point $\mathfrak{r} \in \mathcal{R}_{\mathfrak{p}}$ using the exponential map, such that $\mathfrak{r} = \exp_{\mathfrak{p}}(x^a e_a)$. Here, $\exp_{\mathfrak{p}}: T_{\mathfrak{p}}(\mathcal{M}) \to \mathcal{M}$ is the exponential map at $\mathfrak{p}$, $T_{\mathfrak{p}}(\mathcal{M})$ is the tangent space at $\mathfrak{p}$, and $e_a \in T_{\mathfrak{p}}(\mathcal{M})$ represents the basis vectors. The local rest space of the entire curve $\gamma$ is then defined as $\mathcal{R} \equiv \cup_{\mathfrak{p} \in \gamma} \mathcal{R}_{\mathfrak{p}}$, which provides a local foliation of the spacetime $\mathcal{M}$ around $\gamma$. Any point in $\mathcal{R}$ is described by the Fermi normal coordinates $(\tau, x^1, x^2, x^3)$.

As a result of this construction and the decomposition of the metric along $\gamma$ as $g_{\mu \nu} = - u_{\mu} u_{\nu} + \delta_{i j} e^{i}_{\ \mu} e^{j}_{\ \nu}$, the spatial distance from a point $\mathfrak{r} \in \mathcal{R}$ to the curve $\gamma$ is given by $r^2 = \delta_{i j} x^i x^j$~\cite{Poisson2011}. In addition, the Fermi normal coordinates $(\tau, x^1, x^2, x^3)$ allow us to express the components of the metric around the curve $\gamma$ as
\begin{align}
 & g_{\tau \tau} = - (1 + a_i(\tau) x^i)^2 - R_{\tau i \tau j}(\tau) x^i x^j + \mathcal{O}(r^3), \nonumber \\
 & g_{\tau i} = - \frac{2}{3} R_{\tau jik}(\tau) x^j x^k + \mathcal{O}(r^3), \label{eq:metricfnc}\\
 & g_{i j} = \delta_{ij} - \frac{1}{3} R_{ikjl}(\tau)x^k x^l + \mathcal{O}(r^3), \nonumber
\end{align}
where $a^{\mu}(\tau)$ and $R_{\mu \nu \alpha \beta}(\tau)$ represent, respectively, the $4$-acceleration and components of the Riemann curvature tensor in the Fermi normal coordinates evaluated at the point $\gamma(\tau)$. Finally, the red-shift factor $z(\tau)$, defined by the norm of the $1$-form $\dd \tau$ \cite{Perche2022}, is given by 
\begin{align}
    z(\tau) = \abs{g_{\tau \tau} - g^{ij} g_{\tau i} g_{\tau j}}^{1/2}. \label{eq:redshi}
\end{align}

\subsection{Hamiltonian dynamics of a localized quantum system in curved spacetimes}
\label{sec:iib}

In this section, we construct the Hamiltonian of a non-relativistic quantum particle with some internal degree of freedom in a curved spacetime around the time-like trajectory $\gamma$ of the laboratory frame discussed in the previous section. It is worth mentioning that Ref.~\cite{Perche2022} provides a formal and elegant description of a localized quantum system in a curved spacetime using the Fermi normal coordinates, which we follow closely. However, our construction is also related to the formulation reported in Refs.~\cite{Zych2011,Pikovski2015}. We start by giving the classical description of the Hamiltonian of a particle with some internal structure, which we later quantize.

\begin{figure}
    \centering
    \includegraphics[scale=0.5]{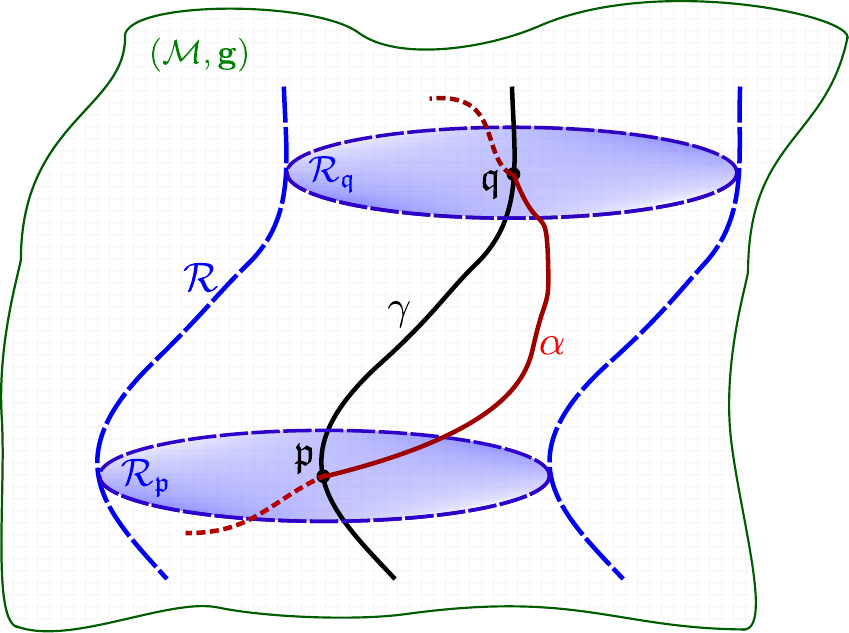}
    \caption{The world-line  $\alpha \subset \mathcal{R}$ of the system and the laboratory world-line $\gamma$.}
    \label{fig:protocol}
\end{figure}

Let us start by considering our quantum particle with some internal structure moving along its worldline, denoted by $\alpha$, near the trajectory $\gamma$ of the laboratory frame. Specifically, the timelike curve $\alpha$ lies within the local rest space $\mathcal{R}$ associated with the curve $\gamma$, where Fermi normal coordinates hold. See the sketch in Fig.~\ref{fig:protocol}. The particle's $4$-momentum along the worldline $\alpha$ is represented by $p^{\mu}$ in this coordinate system. In contrast, in the particle's rest frame, described by primed coordinates $x^{\mu'}$, it is straightforward to show that $p^{j'} =  (\partial x^{j'}/\partial x^{\mu}) p^{\mu} = 0$. This indicates that the total energy, as measured by a commoving observer, is given by  $p_{t'}$ ($x^{0'} \equiv t'$). This total energy includes not only the contribution from the system's rest mass but also any kinetic or binding energies associated with the internal degrees of freedom, which are captured by the particle's internal Hamiltonian $H_{\text{int}}$. Hence
\begin{align}
    p_{t'} = m + H_{\text{int}} \equiv H_{\text{rest}}. \label{eq:Hrest}
\end{align}

On the other hand, $p^{\tau}$ characterizes the motion of the particle relative to the laboratory frame, as described by the Fermi coordinate system. This includes contributions from both the internal and external degrees of freedom. As a result, it represents the total Hamiltonian of the system with respect to the coordinates $(\tau, x^1, x^2 , x^3)$, and we denote it as $H = p_{\tau}$. From $p_{\mu} p^{\mu} = p_{\mu'}p^{\mu'}$, the following relation is obtained: 
\begin{align}
    H = \sqrt{\frac{g^{t't'}H^2_{\text{rest}} - p_j p^j}{g^{\tau \tau}}}.
\end{align}
By selecting the component $x^{t'}$ associated with the commoving observer as the proper time along the particle's world line, it is possible to express the Hamiltonian as
\begin{align}
    H & = \sqrt{-\frac{H^2_{\text{rest}} + p_j p^j}{g^{\tau \tau}}}  = z(\tau) \sqrt{H^2_{\text{rest}} + p_j p^j},
\end{align}
once $g^{t't'} = \eta^{tt} = -1$ and $z(\tau) = (- g^{\tau \tau})^{1/2} = \abs{g^{\tau \tau}}^{1/2}$ are the redshift factors given in Eq.~\eqref{eq:redshi}. 

Now, for a non-relativistic particle such that the internal and kinetic energies are much smaller than its rest energy, we can expand the redshift factor as~\cite{Perche2022}
\begin{align}
    z(\tau) \approx 1 + a_i(\tau) x^i + \frac{1}{2}R_{\tau i \tau j}(\tau) x^i x^j,  \label{eq:redshifapp}
\end{align}
and Hamiltonian as $\sqrt{H^2_{\text{rest}} + p_j p^j} \approx H_{\text{rest}} + \frac{p^2}{2  H_{\text{rest}}}$, where $p^2 \equiv p_j p^j$, which gives us  
\begin{align}
    H(\tau) = & H_{\text{cm}}(\tau) + \mathcal{Z}(\tau) H_{\text{int}} \label{eq:curvhamil}, 
\end{align}
with $\mathcal{Z}(\tau)$ being defined as
\begin{align}
    \mathcal{Z}(\tau) \equiv 1 - \frac{p^2}{2m^2} + a_i(\tau) x^i + \frac{1}{2}R_{\tau i \tau j}(\tau) x^i x^j. \label{eq:timedila}
\end{align}
Moreover, $H_{\text{cm}}$ is the Hamiltonian of the center-of-mass and it is given by
\begin{align}
    H_{\text{cm}}(\tau) = m + \frac{p^2}{2m} + m a_i(\tau) x^i + \frac{m}{2}R_{\tau i \tau j}(\tau) x^i x^j. \label{eq:hamilcm}
\end{align}
In this limit, we can interpret the last two terms of Eq.~\eqref{eq:hamilcm} as perturbations to the (flat space-time) free Hamiltonian $H_0 = m + \frac{p^2}{2m}$. The quantity $\mathcal{Z}(\tau)$ defined by Eq.~\eqref{eq:timedila} can be interpreted as the time dilation factor between the proper time $\tau$ of the laboratory frame and the proper time $t'$ of the system of interest, once
\begin{align}
    \frac{\dd t'}{\dd \tau} & = \sqrt{-g_{\tau \tau} - g_{ij} \frac{\dd x^i}{\dd \tau}\frac{\dd x^j}{\dd \tau}}  \approx 1 - \frac{p^2}{2m^2} + a_i(\tau) x^i + \frac{1}{2}R_{\tau i \tau j}(\tau) x^i x^j = \mathcal{Z}(\tau).
\end{align}

If the particle being considered is a quantum system with internal degrees of freedom, we can apply a quantization procedure to the classical framework by associating a Hilbert space $\mathcal{H_{\text{cm}}} \otimes \mathcal{H}_{\text{int}}$, which represents the composite Hilbert space of the center-of-mass and internal degrees of freedom. The total Hamiltonian $H$, as defined in Eq.~\eqref{eq:curvhamil}, is a Hermitian operator that acts within this composite Hilbert space. Since the redshift factor $z(\tau)$ is a function of the space-like Fermi coordinates, and the time dilation factor $\mathcal{Z}(\tau)$ is a function of the space-like Fermi coordinates and the momentum of the center of mass, the quantization procedure promotes the space dependence of $z(\tau)$ to a function of the position operator and $\mathcal{Z}(\tau)$ to a function of the position and momentum operators, acting on the Hilbert space of the center-of-mass $\mathcal{H}_{\text{cm}}$ and potentially on the Hilbert space of the internal degrees of freedom $\mathcal{H}_{\text{int}}$. For instance, if the internal degrees of freedom are relative position with respect to the center of mass, then it  also depends on the relative position between the world-line of the laboratory frame and the internal degree of freedom and, therefore, $\mathcal{H}_{\text{int}} \simeq L^2(\mathcal{R}_{\tau})$ and $\mathcal{Z}(\tau)$ act on the composite Hilbert space. On the other hand, if the internal degrees of freedom are the spin, then $\mathcal{Z}(\tau)$ will act only in the Hilbert space of the center of mass. 

Using this framework, we can express the Schr\"odinger equation that governs the unitary evolution of our quantum system. Given that the quantum state of our system is represented by $\ket{\psi} \in \mathcal{H}^{(\tau)}_{\text{cm}} \otimes \mathcal{H}_{\text{int}}$, where $\mathcal{H}^{(\tau)}_{\text{cm}} \simeq L^2(\mathcal{R}_{\tau})$ with $L^2(\mathcal{R}_{\tau})$ being the space of square-integrable functions over the rest space $\mathcal{R}_{\tau}$. Here, $\tau$ denotes the proper time along the worldline $\gamma$ of the laboratory frame, which is used to label a specific point $\mathfrak{p}$ on this worldline. Therefore, the Schr\"odinger equation takes the form
\begin{align}
    i \partial_{\tau} \ket{\psi} = H(\tau) \ket{\psi} = \left(H_{\text{cm}}(\tau) + \mathcal{Z}(\tau) H_{\text{int}} \right) \ket{\psi}.\label{eq:schro}
\end{align}
From the above equation, the global unitary evolution operator $U$ can be defined by $U(\tau) \ket{\psi(\tau = 0)} = \ket{\psi(\tau)}$, such that $U:\mathcal{H}^{(0)}_{\text{cm}} \otimes \mathcal{H}_{\text{int}} \to \mathcal{H}^{(\tau)}_{\text{cm}} \otimes \mathcal{H}_{\text{int}}$, also satisfying the Schr\"odinger equation.

\section{Entropy production in a curved spacetime}
\label{sec:III}
In this section, we describe the protocol employed to derive the detailed fluctuation theorem under the two-point measurement scheme for a localized quantum system in a curved spacetime, which was first presented in Ref.~\cite{Basso2024}.

Let us begin by recalling that a spacetime $(\mathcal{M}, \mathbf{g})$ is said to be time-orientable if it is possible to consistently distinguish between future-directed and past-directed timelike vectors throughout the entire manifold~\cite{Wald}. In our case, we only require that at least a portion of the spacetime $(\mathcal{M}, \mathbf{g})$ be time-orientable, specifically the local rest space $\mathcal{R} \subset \mathcal{M}$ associated with the curve $\gamma$. This assumption holds because we consider $\gamma$ to be a timelike curve representing the worldline of the laboratory frame, which is oriented toward the future. Furthermore, from the curve $\gamma$ and its local rest space $\mathcal{R}$, we derived the Hamiltonian~\eqref{eq:curvhamil}, which governs the system's evolution and establishes a concept of time flow in the sense described by Connes and Rovelli~\cite{Connes1994}, which makes it possible to define local thermal equilibrium states~\cite{Basso2023,Basso2024,Connes1994,Rovelli2011,Rovelli2013}.

Moving on, we introduce two protocols: one corresponding to the forward direction of time and the other to the reverse direction. The difference between these two processes will serve as a measure of the irreversibility of the forward process~\cite{Crooks1998}. In both cases, the system is initially prepared in an equilibrium state, its energy is measured, it then evolves under a specific quantum map, and finally, its energy is measured again. From the outcomes of these measurements, we can quantify entropy production.

As illustrated in Fig.~\ref{fig:protocol}, the forward process starts at the point $\mathfrak{p} = \gamma(\tau = 0) \in \mathcal{M}$, in the intersection of the curves $\alpha$ and $\gamma$, where the observer performs the first projective energy measurement. We assume that the state of our system is given by $\rho_0 = \ketbra{x_0} \otimes \sigma_{0}$, with $\ket{x_0}$ representing the state of the external degrees of freedom while the internal degrees of freedom are described by the thermal state $\sigma_{0} = e^{-\beta \mathfrak{h}(0)}/Z_0,$ with $\mathfrak{h}(0) = \mathcal{Z}(0) H_{\text{int}}$ and $Z_{0} = \Tr{e^{-\beta \mathfrak{h}(0)}}$ being the initial Hamiltonian of the internal degrees of freedom and the partition function, respectively. The inverse temperature is represented by $\beta$. The measurement performed in the eigenbasis of $\mathfrak{h}(0)$ results in the eigenvalue $\epsilon_{l}^{0}$ with probability $p_{l} = e^{-\beta\epsilon_{l}^{0}}/Z_{0}$. Right after this measurement, the state is updated to a localized non-relativistic quantum system in curved spacetimes:
a general characterization of particle detector mode $\ket{\Psi_0} = \ket{x_0}\otimes \ket{\epsilon_{l}^{0}}$. 

The second step consists in letting the quantum system move along its worldline $\alpha$, with the evolution dictated by the Hamiltonian~\eqref{eq:curvhamil}, such that the state of the system evolves to
\begin{eqnarray*}
     \ket{\Psi(\tau)} &=& \mathcal{T} e^{-i\int_{\alpha} \left( H_{\text{cm}}(\tau) + \mathcal{Z}(\tau)H_{\text{int}}\right) \dd\tau} \ket{x_0}\otimes \ket{\epsilon_{l}^{0}} \\
     &=& \mathcal{T}_{\text{cm}} e^{-i\int_{\alpha} H_{\text{cm}} (\tau) \dd\tau} \ket{x_0} \otimes \mathcal{T}_{\text{int}} e^{-i\int_{\alpha} \mathfrak{h}(\tau) \dd\tau} \ket{\epsilon_{l}^{0}},
\end{eqnarray*}
where $\mathcal{T} \equiv \mathcal{T}_{\text{cm}} \otimes \mathcal{T}_{\text{int}}$ is the time-ordering operator and $\mathfrak{h}(\tau) = \mathcal{Z}(\tau) H_{\text{int}}$. The last equality follows from the semiclassical approximation, in which the motion of the quantum particle along its worldline is well-defined. This implies that the internal state evolves accordingly to $U \equiv U(\tau) = \mathcal{T}_{\text{int}} e^{-i\int_{\alpha} \mathfrak{h}(\tau) \dd\tau}$.

The third, and final, step is realized when the system intersects again the laboratory frame at the point $\mathfrak{q} = \gamma(\tau = T) \in \mathcal{M}$, where a final projective measurement with respect to the internal Hamiltonian $\mathfrak{h}(T) =\mathcal{Z}(T) H_{\text{int}}$ is done. Hence, from the definition of work as the stochastic variable $W_{k,l} \equiv \epsilon_{k}^{T} - \epsilon_{l}^{0}$, we can construct the work probability distribution density of the forward process as $P_{\text{fwd}}(W) = \sum_{k,l}p_{k,l}\delta\left[W - W_{k,l}\right]$, where $p_{k,l} = p_{l}p_{k|l}$ is the joint probability of obtaining $\epsilon_{l}^{0}$ in the first measurement and $\epsilon_{k}^{T}$ in the second one. It follows that
\begin{align}
    P_{\text{fwd}}(W)= \sum_{j,k} \delta\left(W - (\epsilon_{k}^{T} - \epsilon_{l}^{0}) \right) \frac{e^{- \beta \epsilon_{l}^{0}}}{Z_0} \abs{\bra{\epsilon_{k}^{0}}U\ket{\epsilon_{l}^{0}}}^2.
\end{align}

For the reverse process, let us remember that, given a time orientation in the region $\mathcal{R} \subset \mathcal{M}$, general relativity does not forbid us to define past-directed curves in $\mathcal{R}$~\cite{Wald}. Since the curves $\gamma$ and $\alpha$ parameterized by $\tau$ in Fig~\ref{fig:protocol} are directed towards the future, then we can obtain past-directed curves $\gamma'$ and $\alpha'$ by making $\tau \to - \tau$. Hence, the reverse process starts at the point $\mathfrak{q} = \gamma(T) \in \mathcal{M}$, where a first projective energy measurement is realized with the state of the system given by $\rho_{T} = \Theta \ketbra{x_{T}} \otimes \sigma_{T}\Theta^{\dagger}$. Here, $\Theta$ is the anti-unitary time-reversal operator~\cite{Sakurai}. The internal degrees of freedom are supposed to be in the thermal state $\sigma_{T} = e^{-\beta \mathfrak{h}(\tau)}/Z_{T}$, as usual.

Under the assumption that the Hamiltonian~\eqref{eq:curvhamil} is time-reversal invariant, the time-reversal evolution is then governed by the micro-reversibility principle, i.e.,  $\Tilde{U} \equiv \Tilde{U}(T - \tau) = \Theta (\mathcal{T}_{\text{int}}e^{-i\int_{\alpha} \mathfrak{h}(\tau) \dd \tau})^{\dagger}\Theta^{\dagger}$~\cite{Campisi}. The last step is then realized when the system intersects the laboratory frame at point $\mathfrak{p} = \gamma(0) \in \mathcal{M}$, where the final projective energy measurement with respect to the internal Hamiltonian $\mathfrak{h}(0)$ is done. The work probability distribution density of the reverse process is
\begin{align}
    P_{\text{rev}}(-W)= \sum_{j,k} \delta\left( (\epsilon_{k}^{T} - \epsilon_{l}^{0}) - W \right) \frac{e^{- \beta \epsilon_{k}^{T}}}{Z_{T}} \abs{\bra{\epsilon_{l}^{0}}\Tilde{U}\ket{\epsilon_{k}^{0}}}^2,
\end{align}
which allows us to obtain the Crooks relation in a curved spacetime, i. e.,
\begin{align}
    \frac{P_{\text{fwd}}(W)}{P_{\text{rev}}(-W)} = e^{\beta(W - \Delta F)}. \label{eq:crooks}
\end{align}

Moreover, by integrating Eq.~\eqref{eq:crooks} over the probability distributions, we obtain the Jarzynski equality
\begin{equation}
\expval{e^{-\beta W}}_{\alpha, \gamma} = e^{-\beta\Delta F}, \label{eq:jar_time} 
\end{equation}
where the subscripts above remind us that the joint probability distribution depends on the path $\alpha$ the system follows through spacetime, on the acceleration of the curve $\gamma$, and on the components of the curvature tensor evaluated at the curve $\gamma$. Therefore, Eq.~\eqref{eq:jar_time} represents the extent to which the system deviates from this initial equilibrium state during its journey along its trajectory in curved spacetime.

Let us notice that, if $H_{\text{int}}$ depends on $\tau$, the results given by Eqs.~\eqref{eq:crooks} and ~\eqref{eq:jar_time} still hold, with the addition of contributions from the driven component of the Hamiltonian, which alters the rate of entropy production. Moreover, by neglecting the internal degrees of freedom, we can derive Eq.~\eqref{eq:jar_time} for the center-of-mass degrees of freedom by considering a localized quantum system within the local rest space $\mathcal{R}$ of the curve $\gamma$. Then, the total Hamiltonian is the center-of-mass Hamiltonian given by Eq.~\eqref{eq:hamilcm} and the projective energy measurements are realized with respect to $H_{\text{cm}}(\tau)$. 

\section{The newtonian limit and the equivalence principle}

To provide more concrete results, we now examine the Newtonian limit, which was first derived in Ref.~\cite{Basso2023}. This limit applies to the case of a weak and static gravitational field, like the one near Earth's surface, where the gravitational field can be considered uniform.  The spacetime metric of this scenario can be expressed as
\begin{align}
    \dd s^2 = - (1 + 2gx)\dd t^2 + \dd x^2 + \dd y^2 + \dd z^2
\end{align}
where $g = GM/R^2$ is Earth's gravitational acceleration at the origin of the laboratory frame $(x = 0)$, which is located a distance $R$ from Earth's center.

The equivalence principle states that an observer under uniform acceleration in empty spacetime is equivalent to an observer at rest under the action of a uniform gravitational field. Therefore, in this case, we can consider that the Fermi normal coordinates are just usual Newtonian coordinates $\{\tau = t, x, y, z\}$ such that, from Eq.~\eqref{eq:metricfnc}, all the components of the Riemann curvature tensor vanish and the only nonzero component of the observer' acceleration is given by $a_x = g$.

Now, let us consider that our quantum system is a free semiclassical system with the center of mass having a well defined trajectory in spacetime while its internal degrees of freedom are described quantum mechanically. Moreover, the internal degrees of freedom are a two level quantum system such that $\mathcal{Z}(\tau)$ will act only in the Hilbert space of the center of mass, as e. g., a spin-$1/2$ particle. In this case, the total Hamiltonian of the system is given by Eq.~\eqref{eq:curvhamil}, where the Hamiltonian of the center-of-mass is given by
\begin{align}
    H_{\text{cm}} = m + \frac{p^2}{2m} + 2 m g x,
\end{align}
while the time dilation factor is   
\begin{align}
    \mathcal{Z}= 1 - \frac{p^2}{2m} + 2gx.
\end{align}

Going back to the work protocol discussed in the previous section, given that $\mathfrak{h}(0) = H_{\text{int}}$ is the initial internal Hamiltonian at the point $\mathfrak{p} \in \mathcal{M}$, one defines $H_{\text{int}}\ket{\epsilon_{m}^{0}} = \epsilon_{m}^{0}\ket{\epsilon_{m}^{0}}$. Since the particle is initially at rest in $x = 0 $, we have $\mathcal{Z} = 1$. After the preparation, a projective energy measurement (defined by $H_{\text{int}}$) is performed on the system and, if the eigenvalue $\epsilon_{m}^{0}$ is obtained, the state of the system just after the measurement is given by $\ket{\epsilon_{m}^{0}}$. The external degrees of freedom do not enter here since we are considering the semiclassical approximation. We let the system evolve by following the trajectory $\alpha$ on spacetime. After a certain amount of proper time $\tau$, the second energy measurement is performed, according to the final internal Hamiltonian, resulting in the eigenvalue 
\begin{align}
     \epsilon_{m}^{\tau} = \mathcal{Z} \epsilon_{m}^{0} \approx \Big(1 + 2 g x - \frac{p^2 }{2m^2}\Big)\epsilon_{m}^{0}.
     \label{eq:eigenvalue}
\end{align}
From this, we can define the work as
\begin{equation}
W_{n,m} = \epsilon_{n}^{\tau} - \epsilon_{m}^{0} =  \left(\mathcal{Z}\epsilon_{n}^{0} - \epsilon_{m}^{0}\right)\delta_{n,m}, 
\label{eq:work}
\end{equation}
representing a shift in the energy eigenvalues.

Now, let us consider that our two-level quantum system is such that $\epsilon_{0}^{0} = 0$ and $\epsilon_{1}^{0} = \epsilon > 0$. Then, we can calculate the dissipated work defined by $W_{\text{diss}} = \expval{W} - \Delta F$, where $\expval{W} = \Tr \{\mathfrak{h}(\tau) \rho_{\tau}\} - \Tr \{\mathfrak{h}(0) \rho_{0}\}$ is the average work, $\rho_0 = e^{-\beta \mathfrak{h}(0)}/Z_{0}$ is the initial thermal state, and $\rho_{\tau} = e^{-\beta \mathfrak{h}(\tau)}/Z_{\tau}$ is the final thermal state. The average entropy production associated with the dissipated work can be defined as $\Sigma = \beta W_{\text{diss}}$, which gives us
\begin{align}
    \Sigma  = (\mathcal{Z} - 1)\beta \epsilon + \ln\left(\frac{1 - e^{- \mathcal{Z}\beta \epsilon }}{1 - e^{-\beta \epsilon}} \right). \label{eq:Sprod}
\end{align}

It is important to observe here that the entropy production $\Sigma$ can be negative, positive, or zero, depending on whether $\mathcal{Z}$ is less than, greater than, or equal to one, respectively. Of course, the sign of $\Sigma$ has the same sign as the dissipated work $W_{\text{diss}}$, which depends on whether the quantum system is doing work on the spacetime or if spacetime is doing work on the quantum system. Moreover, if we consider the system at rest together with the laboratory frame, we obtain $\mathcal{Z} = 1$ and, consequently, $\Sigma = \beta W_{\text{diss}}= 0$. For this case, we do not expect any entropy to be produced since there will be no measurable time dilation.

On the other hand, if $\mathcal{Z} > 1$, then $\Sigma = \beta W_{\text{diss}} > 0$. In our example, this occurs in the limit of large $m$, where the kinematic degrees of freedom become negligible, resulting in $\mathcal{Z} = 1 + 2gx$. This is similar to the well-known fact that a photon must expend energy to overcome the gravitational (metric) potential.

The last possibility happens for trajectories such that $\mathcal{Z} < 1$, implying that $\Sigma = \beta W_{\text{diss}} < 0$. Note that, in this case, both $\expval{W}_{\alpha, \gamma}$ and $\Delta F_{\alpha, \gamma}$ are negative quantities. However, the energy flux can be defined as positive regardless of the direction, whether it's from the system to the field or from the field to the system. As a result, we arrive at the same conclusion when accounting for the redshift scenario.

\section{The expanding universe}
\label{sec:iv}

\begin{figure}
    \centering
    \includegraphics[scale=0.5]{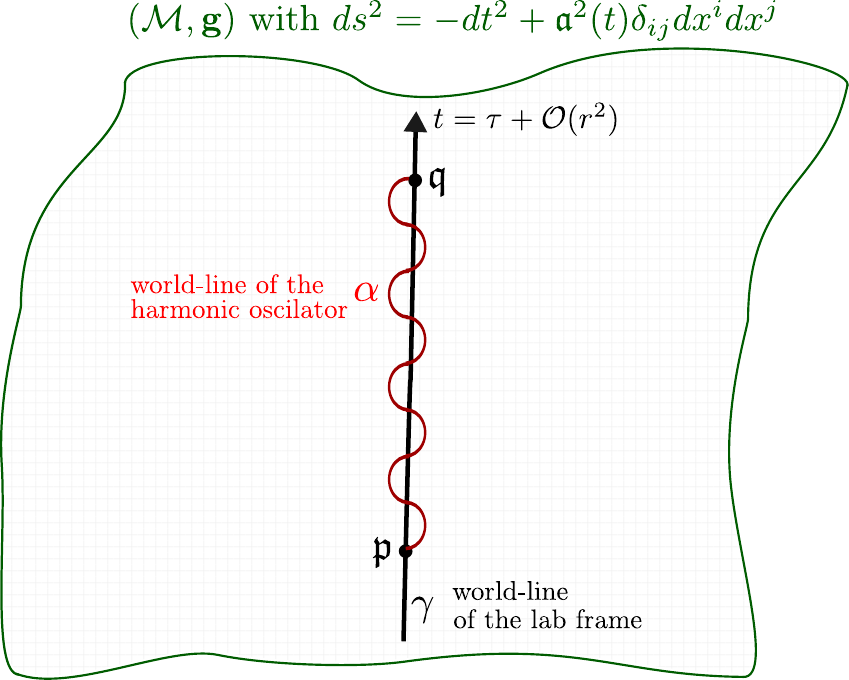}
    \caption{Harmonic oscillator in an expanding universe.}
    \label{fig:universe}
\end{figure}

The second example that we discuss is a quantum mechanical harmonic oscillator (QHO) in an expanding universe as depicted in Fig.~\ref{fig:universe}. This example was selected to demonstrate that our protocol applies to the center-of-mass degrees of freedom while neglecting the internal ones.

An isotropic and homogeneous expanding universe with zero spatial curvature is described by the following metric
\begin{align}
    \dd s^2 = -\dd t^2 + \mathfrak{a}^2(t)\Big(\dd X^2 + \dd Y^2 + \dd Z^2\Big),
\end{align}
in the Friedmann-Robertson-Walker (FRW) coordinates, where $\mathfrak{a}(t)$ is the scale factor. The worldline of our laboratory frame is the geodesic $\gamma$ described by $\tau = \lambda$ and $X = Y = Z = 0$. This implies that the acceleration of our laboratory frame is zero. The relation between the FRW coordinates $\{t, X, Y, Z\}$ and the Fermi normal coordinates $\{\tau, x , y, z\}$ is given by \cite{Cooperstock1998}
\begin{align}
    & t = \tau - \frac{\dot{\mathfrak{a}}}{2 \mathfrak{a}}r^2 + \mathcal{O}(r^4), \ \ \ X^i = \frac{x^i}{\mathfrak{a}} \Big(1 + \frac{\dot{\mathfrak{a}}^2}{3 \mathfrak{a}^2} r^2\Big) + \mathcal{O}(r^4), 
\end{align}
where $r \equiv \delta_{ij}x^i x^j$ and $\dot{\mathfrak{a}} = \dd \mathfrak{a}/ \dd t$. The transformation above allows us to rewrite the metric in terms of the Fermi normal coordinates, i.e.,
\begin{align}
    \dd s^2 = -\Big(1 - \frac{\Ddot{\mathfrak{a}}}{\mathfrak{a}}r^2\Big)\dd\tau^2 + \Big[\delta_{ij} - \frac{\dot{\mathfrak{a}}^2}{\mathfrak{a}^2}\Big(\frac{r^2 \delta_{ij} - x_i x_j}{3} \Big) \Big] \dd x^i \dd x^j,
\end{align}
which implies that the components of the curvature in the Fermi normal coordinates are given by
\begin{align}
    & R_{\tau x \tau x} = R_{\tau y \tau y} = R_{\tau z \tau z} = - \frac{\Ddot{\mathfrak{a}}}{\mathfrak{a}}, \\
    & R_{xyxy} = R_{xzxz} = R_{yzyz} = \frac{\dot{\mathfrak{a}}^2}{\mathfrak{a}^2}.
\end{align}

Given that the stage is set, our system is a one-dimensional quantum harmonic oscillator, where the energy eigenvalues are given by $\epsilon_n^0 = (n + 1/2)\omega$ for the unperturbed Hamiltonian $H_0 = \frac{p^2}{2m} + \frac{1}{2}m\omega^2x^2$, which is in thermal equilibrium at a point $\mathfrak{p} = \gamma(0)$ of the worldline of the laboratory frame. At this point, a projective measurement is performed in the eigenbasis $\ket{\epsilon^0_m}$ of $H_0$. The probability of measuring the eigenvalue $\epsilon^0_m$ is $p_m = e^{-\beta \epsilon^0_m}/Z_0$, where $Z_0 = 2/\sinh(\frac{1}{2}\beta\omega)$. 

The next step involves allowing the quantum system to evolve under the following Hamiltonian
\begin{align}
    H(\tau) = H_0 + \frac{1}{2}m R_{\tau x \tau x}(\tau) x^2 = H_0 - \frac{m \Ddot{\mathfrak{a}}}{2 \mathfrak{a}} x^2.  \label{eq:Huni}
\end{align}
From the equation above, we can see that the curvature of the spacetime makes the frequencies time-dependent, i.e.,
\begin{align}
\omega(\tau) = \sqrt{\omega^2_0 - \frac{\Ddot{\mathfrak{a}}}{\mathfrak{a}}}. \label{eq:freq}    
\end{align}
Therefore, the energy of this oscillator is not conserved, which leads to non-zero transition probabilities between the initial and final energy states, as we shall see.

In the interaction picture, the evolution of our system can be described by
\begin{align}
    i \partial_{\tau} \ket{\psi}_I = V_I(\tau) \ket{\psi}_I,
\end{align}
with $V_I(\tau) = e^{i H_0 \tau} V(\tau) e^{-i H_0 \tau}$, $V(\tau) = \frac{1}{2}m R_{\tau x \tau x}(\tau) x^2 $, and $\ket{\psi}_I = e^{i H_0 \tau} \ket{\psi}$. By expanding $\ket{\psi}_I = \sum_n c_n(\tau) \ket{\epsilon_n^0}$, we have
\begin{align}
    c_n(\tau) = - i \sum_m \int^{\tau}_0  \bra{\epsilon^0_n}V(\tau')\ket{\epsilon^0_m} e^{i(n - m) \tau'} c_m(\tau') \dd\tau'.
\end{align}
Since the initial state of our system is $\ket{\epsilon^0_m}$, we have
\begin{align}
c_n(\tau) = - \frac{i m }{2} \bra{\epsilon^0_n}x^2\ket{\epsilon^0_m} f(\tau),   
\end{align}
Since $x = \sqrt{\frac{1}{2m\omega}}(A + A^{\dagger})$ with $A$ and $A^{\dagger}$ being the annihilation and the creation operators, we have 
\begin{align}
  \bra{\epsilon^0_n}x^2\ket{\epsilon^0_m}  = \frac{1}{2m\omega}  \Big(\sqrt{m(m-1)}\delta_{n, m-2} + (2m +1) \delta_{n,m}  +  \sqrt{(m+1)(m+2)}\delta_{n, m+2}\Big),
\end{align}
and 
\begin{align}
    f(\tau) = \int_{0}^{\tau} R_{\tau x \tau x} (\tau') e^{i(k-l) \omega_0 \tau' } \dd \tau' = -  \int_{0}^{\tau} \frac{\Ddot{\mathfrak{a}}}{\mathfrak{a}} e^{i(k-l) \omega_0 \tau' } \dd \tau'. 
\end{align}
From the equations provided, it is evident that the transition probability, $p^{\tau}_{n|m} = \abs{c_{n \neq m}(\tau)}^2$, is influenced by the curvature of the expanding universe, which leads to the generation of entropy, as it is directly associated with the changes in population distributions.

In the universe dominated only by a positive cosmological constant $\Lambda$, i. e., the de Sitter spacetime, the Einstein field equation can be rewritten $G_{\mu \nu} = - \Lambda g_{\mu \nu}$, which corresponds to the energy-momentum tensor of a perfect fluid such that $p_{\Lambda} = - \rho_{\Lambda} = - \frac{\Lambda}{8 \pi}$, where $p_{\Lambda}$ is the isotropic pressure and $\rho_{\Lambda}$ is the positive energy density. In this case, the scale factor is given by $\mathfrak{a}(t) = e^{\mathbb{H} t}$, where $\mathbb{H} = \frac{\dot{\mathfrak{a}}}{\mathfrak{a}} = \sqrt{\Lambda/3}$ is the Hubble parameter and we have $\frac{\Ddot{\mathfrak{a}}}{\mathfrak{a}} = \mathbb{H}^2$. Then, the transition probability for $n \neq m$ is given by
\begin{align}
    p^{\tau}_{n|m} =  4 \Big(\frac{m \mathbb{H}^2}{2}\Big)^2 \frac{\abs{\bra{\epsilon^0_n}x^2\ket{\epsilon^0_m}}^2}{\abs{\epsilon^0_n - \epsilon^0_m}^2} \sin^2\Big(\frac{(n - m)\omega t}{2}\Big). \label{eq:tpho}
\end{align}
Thus, we observe that, in this scenario, the transition probability explicitly depends on the Hubble parameter or the cosmological constant, leading to the generation of entropy. For example, if we initially consider the system in its ground state, the only permissible transition is to the second excited state. A direct calculation reveals that Eq.~\eqref{eq:tpho} gives us $p^{\tau}_{2|0} = (\mathbb{H}/\sqrt{2}\omega_0)^4 \sin \omega_0 t$. Given that $\mathbb{H} \approx 10^{-61} t_p^{-1}$, where $t_p = 5.391 \times 10^{-44}s$ is the Planck time, and $\omega_0 \approx 10^{-30} t_p^{-1}$ ($\omega_0 \approx 10^{13}s^{-1}$) for typical molecular vibrational modes, the ratio $\mathbb{H}/\omega_0$ is of order $10^{-31}$. This indicates that while the transition probability is very small, it is nonetheless non-zero.

\section{Concluding remarks}

This chapter discussed the detailed fluctuation theorem for a localized quantum system living in a general curved spacetime, which reveals how the spacetime curvature can produce entropy.

In order to better understand the role of entropy production due to the curvature of spacetime, let us resort to the gravito-electromagnetic analogy discussed, for instance, in Refs.~\cite{Costa2014,Ruggiero2021} and define the gravito-electric potential as $\phi(\tau) \equiv - \frac{1}{2} R_{\tau i \tau j}(\tau) x^i x^j,$ such that the gravito-electric field (up to linear order in $x^i$) is given by $E_i(\tau) = R_{\tau i \tau j}(\tau) x^j$. The contribution of the gravito-magnetic potential in the gravito-electric field is second order and therefore will not be considered in our analysis. Thus, we can describe the term $\frac{m}{2}R_{\tau i \tau j}(\tau) x^i x^j$ that appears in Eq.~\eqref{eq:hamilcm} as $m E_i(\tau) x^i$, while the term $\frac{1}{2}R_{\tau i \tau j}(\tau) x^i x^j H_{int}$ that appears in $\mathcal{Z}(\tau) H_{\text{int}}$ can be written as $H_{\text{int}} E_i(\tau) x^i$. It is noteworthy the similarity of these two terms with the electric dipole interaction, with both $m$ and $H_{int}$ playing the role of the charge of the gravitational field, which is reasonable since (internal) energy also gravitates in general relativity. Therefore, we can interpret the terms $m E_i(\tau) x^i$ and $ H_{\text{int}} E_i(\tau) x^i$ as the gravitational analogue of a charged quantum system interacting with a time-dependent electric field. 

Our main result implies that entropy production is not an invariant quantity defined solely by the system. Rather, it depends on the observer who measures it, since it depends on the worldline of the laboratory in an arbitrary spacetime. This is a robust result that goes in the same direction as those discussed in Refs.~\cite{Wald01,Marolf} regarding the subtleties of defining entropy in a curved spacetime. Specifically, two different families of observers will not agree on the entropy production in general. It is worth remembering that, for comparison, each family of observers has to realize the same protocol, since the measurements in the energy basis are locally performed.

Additionally, our findings establish a deep and fundamental link between the time-orientability of the laboratory frame's worldline $\gamma$ and the production of entropy and, therefore, with the thermodynamic arrow of time. This is the precise meaning of the observer-dependent nature of entropy production. Such orientability is needed in order to obtain the Hamiltonian~\eqref{eq:curvhamil}, which governs the evolution of the quantum system and thereby defines a notion of time flow and thermal equilibrium reference states, as discussed in Refs.~\cite{Connes1994,Rovelli2011,Rovelli2013}. For instance, in the quantum harmonic oscillator in an expanding universe, the curvature drives the quantum system out of equilibrium due to the last term in Eq.~\eqref{eq:Huni}, causing the change in the populations. Therefore, our result shows that the arrow of time is rooted in the causal structure of spacetime.

\section*{Acknowledgments}

This work was supported by the National Institute for the Science and Technology of Quantum Information (INCT-IQ), Grant No.~465469/2014-0, by the National Council for Scientific and Technological Development (CNPq), Grants No.~308065/2022-0, and by the Coordination of Superior Level Staff Improvement (CAPES).

\renewcommand{\bibname}{References}
\begingroup
\let\cleardoublepage\relax

\endgroup

\end{document}